\documentclass{article}
\usepackage{spconf,amsmath,graphicx}

\ninept

\title{Improved acoustic-to-articulatory inversion using representations from pretrained self-supervised learning models}

\name{Sathvik Udupa $^1$ , Siddarth C $^2$ \thanks{$^2$ Work performed before the author joined the Indian Institute of Technology Madras}, Prasanta Kumar Ghosh $^1$}
\address{
  $^1$ Electrical Engineering Department, Indian Institute of Science (IISc), Bangalore - 560012, India.\\
  $^2$ Robert Bosch Centre for Data Science
and AI, Indian Institute of Technology Madras - 600036, India.
}

\begin{document}
\maketitle
\begin{abstract}
In this work, we investigate the effectiveness of pretrained Self-Supervised Learning (SSL) features for learning the mapping for acoustic to articulatory inversion (AAI). Signal processing-based acoustic features such as MFCCs have been predominantly used for the AAI task with deep neural networks. With SSL features working well for various other speech tasks such as speech recognition, emotion classification, etc., we experiment with its efficacy for AAI. We train on SSL features with transformer neural networks-based AAI models of 3 different model complexities and compare its performance with MFCCs in subject-specific (SS), pooled and fine-tuned (FT) configurations with data from 10 subjects, and evaluate with correlation coefficient (CC) score on the unseen sentence test set. We find that acoustic feature reconstruction objective-based SSL features such as TERA and DeCoAR work well for AAI, with SS CCs of these SSL features reaching close to the best FT CCs of MFCC. We also find the results consistent across different model sizes.
\end{abstract}


\section{Introduction}
\label{sec:intro}

Acoustic to Articulatory Inversion (AAI), also known as speech inversion, refers to the task of estimating the articulators' trajectories given the corresponding acoustic signal. AAI, being a problem of one-to-many mapping targeting highly dynamic trajectories at high frequency, is very difficult to solve. Identical speech signals could be produced by a variety of vocal tract configurations \cite{Lindblom1979FormantFO}. To address this one-to-many mapping, it is essential to use a well-representative feature set and a paradigm capable of effectively exploiting the inherent temporal relationships.

The literature on AAI is filled with various approaches to model the inversion task. The very first is code book-based methods \cite{Larar1988VectorQO, Atal1978InversionOA}. Later methods adopted Gaussian Mixture Models \cite{Richmond2001MixtureDN, Toda2004AcoustictoarticulatoryIM} and Hidden Markov Models \cite{youssef2009acoustic, zhang2008acoustic}. These approaches were replaced by Recurrent and Convolutional Neural Network modelling \cite{Bozorg2020AcoustictoArticulatoryIW, Liu2015ADR, Illa2017ACS, aravind_blstm, aravind_cnn}, which achieved many significant performances. Neural networks were able to extract the temporal relationships between different input frames and were also capable of generalizing across unseen samples. The state-of-the-art AAI performance is achieved by Transformer \cite{transformer}-based models \cite{udupa_aai_is21, udupa_aai_is22}.

The model's feature set is equally crucial as the model itself. These inputs could be broadly classified into two categories a) acoustic representation and b) phonetic representation. The majority of the past literature uses Mel-features to represent the acoustic signal. Mel frequency cepstral coefficients (MFCCs) are one of the widely used Mel features in AAI and are accepted to be the optimal representation for AAI \cite{Ghosh2010AGS}. In comparison, very few have focused on using phonemes to predict the trajectories \cite{udupa_aai_is21, Zhu2015ArticulatoryMP}. Combining these two features has also been found to generate more accurate trajectories \cite{abhay_tphn}. A few other works try to combine other modalities like real-time MRI and speaker vectors \cite{illa2020speaker} to improve performance.

Recently, many spoken language understanding tasks have started to adopt parameterized representations of speech, rather than the raw waveform or Mel-features. These parameterized features are often learnt via Self-Supervised Learning (SSL) and found to perform better in downstream tasks. SSL has been employed in ASR \cite{wav2vec, vq_wav2vec}, sentiment analysis \cite{Mockingjay, tera} and various other tasks \cite{Mockingjay, tera}. It generally involves training a model on a huge unlabelled corpus and then using the model to extract features to address a downstream task.

We look at understanding what type of SSL features could assist AAI. Based on the nature of training samples, SSL methods could be categorized into 2 classes, predictive and contrastive. The objective of predictive SSL is to maximize the similarity between the representations of two views of the same object without taking into account how closely they resemble the negative samples. Whereas contrastive methods aim to distinguish samples from 2 different objects, learning better distinctive features. While doing so, it is found that contrastive SSL features represent the difference between samples well, but fail to encode the complete information. In general, the former is found to perform better and the latter is preferred when the pretext and downstream tasks are closely related \cite{Liu2022AudioSL}. 

SSL methods have been utilised effectively for a range of tasks involving speech input. However, to our knowledge, there has been very little, if any, work on SSL for speech inversion. Previously, two pretrained features were investigated for 3D tongue modelling \cite{Medina_2022_CVPR}, and MFCC were shown to outperform one of the pretrained features in terms of objective measures. Furthermore, the models have a substantially higher parameter count and are trained on small fixed input windows.

In this study, we explore if AAI could benefit from the pre-trained SSL models. We employ a variety of SSL models, namely APC \cite{apc}, NPC \cite{npc}, PASE+ \cite{pase}, wav2vec \cite{wav2vec}, vq\_wav2vec \cite{vq_wav2vec}, Mockingjay \cite{Mockingjay}, AudioALBERT \cite{audio_albert} and TERA \cite{tera}. Each of the aforementioned models utilizes different training methods, resulting in weights that represent different features, of varying dimensions. The extracted features are then passed on to transformer models to estimate the articulatory trajectories. To have a fair evaluation across input SSL features of different dimensions, we use three different configurations of the AAI transformer with the number of parameters ranging from 2.1M to 15M.


\section{Dataset}
\label{sec:format}

We use an AAI corpus consisting of concurrent acoustic and articulatory data from a collection of 460 phonetically balanced English sentences from the MOCHA-TIMIT corpus \cite{Mocha460}. These sentences were articulated by 10 native Indians, 4 female and 6 male, of ages ranging from 20 to 28 years. All subjects speak fluent English and have had no speech difficulties in the past. The subjects were familiarized with all the sentences to eliminate elocution issues. 

The audio and articulatory trajectories were simultaneously recorded using an Electromagnetic Articulograph AG501 \cite{AG501}. Six sensors, glued on different articulators in accordance to \cite{Zhu2015ArticulatoryMP}, recorded the movement at 250Hz. The six locations are Upper Lip (UL), Lower Lip (LP), Jaw, Tongue Tip (TT), Tongue Body (TB), and Tongue Dorsum (TD). In this work, we consider only the motions in the midsagittal plane which corresponds to the horizontal and vertical movement of the sensors (x and y axis). Thus each audio sample corresponds to 12 trajectories (= 6 sensors $\times$ 2 axes).

\section{Proposed Methodology}
We experiment with various SSL pretrained features as an alternative to MFCCs. In this section, we describe the model architecture used.

We use a standard non-autoregressive transformer model\cite{udupa_aai_is21} for modelling the AAI task. Transformers have been shown to perform well in different configurations of AAI \cite{udupa_aai_is21, udupa_aai_is22} in the past. 
Since the SSL features vary in their feature dimension, we use the transformer AAI in three different configurations to have a fair assessment of results across models. Thus, we have AAI-s (2.1M), AAI-m (7.5M), and AAI-l (15M) representing small, medium and large models respectively, and the number of parameters is reported in brackets (in millions). The model size is controlled by the number of transformer layers and feature dimensions. All the models have a single self-attention head, we did not find any benefit in using multiple attention heads in each layer. The hyperparameters used are shared in the GitHub repository.

\begin{figure}[h]
    \centering
    \includegraphics[width=8cm]{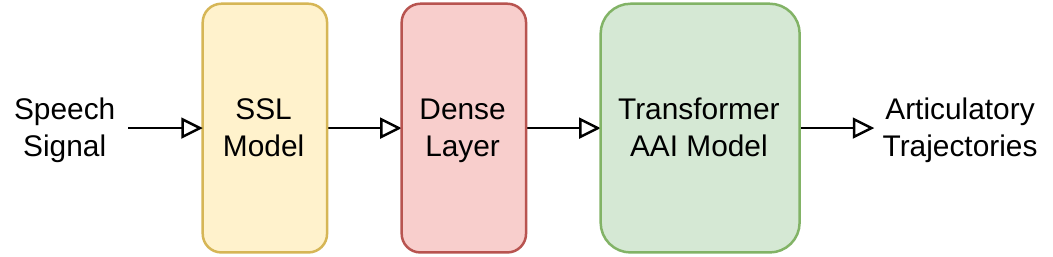}
    \caption{Illustration of how pretrained features are used. Since various pretrained features are of different dimensions, they are first projected to a fixed dimension, before passing it to the transformer. Note that, we only utilise the features from the pretrained SSL model, the SSL model weights are not updated.}
    \label{fig:my_label}
\end{figure}

\section{Experimental Setup}

In this section, we go over the details of various pretrained features used, their training methodology and how they differ. Later on, we summarise the experiment details.

\subsection{Pretrained features}
We use features from self-supervised learning (SSL)  methods on speech. These features are learnt over various formulations of objective functions such as predictive and contrastive loss. They are also learnt over various features such as waveform, Mel spectrogram etc. We briefly go over the various methods used below and summarise the comparison in table \ref{tab:ssl_feats}. For all features, we take the last layer output as the SSL features since we found them to do optimally. All features are taken from the upstream model weights from s3prl \footnote{https://github.com/s3prl/s3prl}.

\textbf{APC:} Autoregressive Predictive Coding \cite{apc} is one of pioneer works on audio SSL. Inspired by large language models, the APC aims to predict the next token autoregressively. APC is a unidirectional model and uses an LSTM to predict the next audio representation.

\textbf{NPC:} NPC \cite{npc} is a non autoregressive method that learns representation conditioned on the local dependency. It is conditioned on a fixed-length window of local log Mel-spectrogram features and estimates the middle frame, which is masked. This allows faster training and inference time and is unrestricted in terms of sequence length, which is crucial for streaming-based downstream tasks.

\textbf{PASE+:} PASE+ \cite{pase} uses a combination of  recurrent and convolutional networks to learn the latent space via 12 self-supervised tasks. PASE+ shares a common encoder, a stack of SincNet, ConvNets and QRNN, and small task-specific neural networks called workers.
PASE+ was then employed on ASR to get significant improvements over standard acoustical features.

\textbf{wav2vec:} wav2vec \cite{wav2vec} was one of the foremost models that employed SSL to address ASR. With raw waveforms as input, wav2vec was trained using a contrastive loss. This allows the model to have an improved understanding of the speech data, leading to better speech representations, and ultimately producing the state-of-the-art ASR model.

\begin{table}
\centering
\caption{This table summarises the differences between various SSL features. \\}
\label{tab:ssl_feats}
\begin{tabular}{|c|c|c|c|c|} 
\hline
\textbf{Feature}     & \textbf{dim } & \textbf{objective} & \textbf{input} & \textbf{prediction}  \\ 
\hline
\textbf{MFCC}        & 13            & -                  & waveform       & -                    \\ 
\hline
\textbf{PASE+}       & 256           & MSE, BCE           & waveform       & multi-task           \\ 
\hline
\textbf{vq\_wav2vec} & 512           & Contrastive        & waveform       & future timesteps     \\ 
\hline
\textbf{wav2vec}     & 512           & Contrastive        & waveform       & future timesteps     \\ 
\hline
\textbf{TERA}        & 768           & L1                 & log melspec    & masked/distort       \\ 
\hline
\textbf{AALBERT}     & 768           & L1                 & log melspec    & masked               \\ 
\hline
\textbf{Mockingjay}  & 768           & L1                 & melspec        & masked               \\ 
\hline
\textbf{APC}         & 512           & L1                 & log melspec    & future timesteps     \\ 
\hline
\textbf{NPC}         & 512           & L1                 & log melspec    & masked               \\ 
\hline
\textbf{DeCoAR}      & 2048          & L1                 & MFCCs          & Reconstruction       \\
\hline
\end{tabular}
\end{table}

\def\arraystretch{1.15}
\begin{table*}

\centering
\caption{This table represents the correlation coefficient (CC) for x and y axis of each articulator, in SS, pooled and FT setup for MFCC and TERA input features, for AAI-m model, averaged across test set utterances of all subjects.\\}
\setlength{\tabcolsep}{0.4em} 
\label{tab:art_spec}
\begin{tabular}{|c|c|c|c|c|c|c|c|c|c|c|c|c|c|c|} 
\hline
setup  & feature & ULx                         & ULy                         & LLx                         & LLy                         & Jawx                        & Jawy                        & TTx                       & TTy                         & TDx                         & TDy                         & TBx    & TBy    & mean (std)                          \\ 
\hline
SS     & MFCC    & 0.7345                      & 0.6873                      & 0.7746                      & 0.8223                      & 0.8289                      & 0.8184                      & 0.8636                    & 0.8652                      & 0.8739                      & 0.8694                      & 0.8734 & 0.8654 & \multicolumn{1}{l|}{0.8231(0.061)}  \\ 
\cline{2-15}
       & TERA    & 0.7751                      & 0.7554                      & 0.8155                      & 0.8794                      & 0.868                       & 0.8622                      & 0.8954                    & 0.9074                      & 0.9067                      & 0.9034                      & 0.9049 & 0.9007 & 0.8645(0.054)                       \\ 
\hline
pooled & MFCC    & 0.7753                      & 0.7366                      & 0.8101                      & 0.8688                      & 0.8625                      & 0.8498                      & 0.901                     & 0.909                       & 0.9113                      & 0.8986                      & 0.9088 & 0.8952 & \multicolumn{1}{l|}{0.8606(0.058)}  \\ 
\cline{2-15}
       & TERA    & 0.8029                      & 0.7813                      & 0.84                        & 0.8952                      & 0.886                       & 0.8761                      & 0.9167                    & 0.9311                      & 0.926                       & 0.9192                      & 0.9244 & 0.9147 & 0.8845(0.051)                       \\ 
\hline
FT     & MFCC    & 0.7753                      & 0.7366                      & 0.8101                      & 0.8688                      & 0.8625                      & 0.8498                      & 0.901                     & 0.909                       & 0.9113                      & 0.8986                      & 0.9088 & 0.8952 & \multicolumn{1}{l|}{0.8606(0.058)}  \\ 
\cline{2-15}
       & TERA    & \multicolumn{1}{r|}{0.8024} & \multicolumn{1}{r|}{0.7862} & \multicolumn{1}{r|}{0.8417} & \multicolumn{1}{r|}{0.9006} & \multicolumn{1}{r|}{0.8897} & \multicolumn{1}{r|}{0.8813} & \multicolumn{1}{r|}{0.92} & \multicolumn{1}{r|}{0.9345} & \multicolumn{1}{r|}{0.9296} & \multicolumn{1}{r|}{0.9224} & 0.9279 & 0.9179 & 0.8879(0.051)                       \\
\hline
\end{tabular}
\end{table*}
\textbf{vq\_wav2vec:} Combining the best of both worlds of NLP and ASR, vq\_wav2vec \cite{vq_wav2vec} achieved the state of the art on the WSJ and TIMIT benchmarks. vq\_wav2vec employs wav2vec's contrastive loss to learn a quantized representation of fixed-length audio samples. Further, these representations are fed into a BERT to learn richer representations, which are finally decoded by an acoustic model to transcribe the audio.

\textbf{Mockingjay:} Unlike the previous uni-directional methods that condition only on the past frames, Mockingjay \cite{Mockingjay} adopts a bidirectional approach. Unlike \cite{vq_wav2vec}, Mockingjay doesn't quantize speech but rather uses a modified BERT to learn representations from Mel-features. These learned representations were found to outperform other representations and features in downstream tasks, such as phoneme classification and sentiment analysis.

\textbf{DeCoAR:}
Similar to Mockingjay, DeCoAR \cite{decoar} uses a bidirectional approach. DeCoAR employs Bi-LSTMs, conditioned on MFCCs, to predict slices of missing MFCC frames. The trained DeCoAR model was used for ASR to achieve significant improvements over standard acoustical features.

\textbf{AALBERT:} AABLERT \cite{audio_albert} or Audio Albert is a frugal version of Mockingjay that has 91\% lesser parameters but achieves results comparable with Mockingjay. AALBERT achieves this by weight tying i.e., sharing parameters across various layers. Similar to Mockingjay, AALBERT showed impressive performance on downstream tasks. 

\textbf{TERA:} Unlike past attempts, TERA \cite{tera} learns representations from altered data in different axes, namely time frequency and magnitude. While past methods that used contrastive loss were used on raw data, TERA was trained on Mel features. It was also found that, while larger models are desirable for fine-tuning, smaller models are better for extracting features.

Table \ref{tab:ssl_feats} summarises the differences between various SSL features. The features have different loss functions based on their learning paradigm. The inputs to the SSL models differ as well ranging from waveform to different acoustic features such as log Mel spectrogram, MFCC, and Mel spectrogram. The prediction task is also different across SSL models. While PASE+ is a multi-task prediction model, few models predict features of future time steps and few others learn by predicting a masked part of the input, and this includes distorting the input features w.r.t TERA. Understanding the differences between the SSL features is important so that we can infer what type of objective function and prediction task is appropriate for AAI.

\begin{table*}
\centering
\caption{The correlation coefficient (CC) is reported for subject-specific (SS), pooled and fine-tuned (FT) training across 3 models - AAI-s, AAI-m, AAI-l for different input features. The values are averaged across all articulators, utterances (and subjects in cased of SS and FT) and the standard deviation is reported in brackets. Among input features, MFCC is the baseline acoustic feature and other features are acquired from pretrained SSL models. The best CC for each column is highlighted in bold.}

\setlength{\tabcolsep}{1em} 
\label{tab:full_res}
\begin{tabular}{|c|c|c|c|c|c|c|c|c|c|} 
\hline
Model          & \multicolumn{3}{c|}{AAI-s}                                                                                                                                                                                       & \multicolumn{3}{c|}{AAI-m}                                                                                                                                                                                         & \multicolumn{3}{c|}{AAI-l}                                                                                                                                                                                         \\ 
\hline
Feature        & SS                                                                        & pooled                                                                    & FT                                                       & SS                                                                         & pooled                                                                     & FT                                                       & SS                                                                         & pooled                                                   & FT                                                                         \\ 
\hline
MFCC           & \begin{tabular}[c]{@{}c@{}}0.8059\\ (0.03)\end{tabular}                   & \begin{tabular}[c]{@{}c@{}}0.8435\\ (0.06)\end{tabular}                   & \begin{tabular}[c]{@{}c@{}}0.8724\\ (0.023)\end{tabular} & \begin{tabular}[c]{@{}c@{}}0.8231\\ (0.029)\end{tabular}                   & \begin{tabular}[c]{@{}c@{}}0.8606\\ (0.055)\end{tabular}                   & \begin{tabular}[c]{@{}c@{}}0.8663\\ (0.024)\end{tabular} & \begin{tabular}[c]{@{}c@{}}0.8285\\ (0.026)\end{tabular}                   & \begin{tabular}[c]{@{}c@{}}0.8686\\ (0.055)\end{tabular} & \begin{tabular}[c]{@{}c@{}}0.8517\\ (0.024)\end{tabular}                   \\ 
\hline
PASE\textbf{+} & \begin{tabular}[c]{@{}c@{}}0.8205\\ (0.022)\end{tabular}                  & \begin{tabular}[c]{@{}c@{}}0.8426\\ (0.058)\end{tabular}                  & \begin{tabular}[c]{@{}c@{}}0.8527\\ (0.022)\end{tabular} & \begin{tabular}[c]{@{}c@{}}0.8352\\ (0.022)\end{tabular}                   & \begin{tabular}[c]{@{}c@{}}0.8588\\ (0.054)\end{tabular}                   & \begin{tabular}[c]{@{}c@{}}0.8656\\ (0.021)\end{tabular} & \begin{tabular}[c]{@{}c@{}}0.842\\ (0.021)\end{tabular}                    & \begin{tabular}[c]{@{}c@{}}0.867\\ (0.054)\end{tabular}  & \begin{tabular}[c]{@{}c@{}}0.8741\\ (0.024)\end{tabular}                   \\ 
\hline
vq\_wav2vec    & \begin{tabular}[c]{@{}c@{}}0.7871\\ (0.077)\end{tabular}                  & \begin{tabular}[c]{@{}c@{}}0.7993\\ (0.05)\end{tabular}                   & \begin{tabular}[c]{@{}c@{}}0.8159\\ (0.077)\end{tabular} & \begin{tabular}[c]{@{}c@{}}0.7858\\ (0.086)\end{tabular}                   & \begin{tabular}[c]{@{}c@{}}0.8046\\ (0.048)\end{tabular}                   & \begin{tabular}[c]{@{}c@{}}0.819\\ (0.076)\end{tabular}  & \begin{tabular}[c]{@{}c@{}}0.7846\\ (0.085)\end{tabular}                   & \begin{tabular}[c]{@{}c@{}}0.8097\\ (0.047)\end{tabular} & \begin{tabular}[c]{@{}c@{}}0.8159\\ (0.077)\end{tabular}                   \\ 
\hline
TERA           & \begin{tabular}[c]{@{}c@{}}\textbf{0.8611}\\(0.02)\end{tabular} & \begin{tabular}[c]{@{}c@{}}\textbf{0.8767}\\(0.05)\end{tabular} & \begin{tabular}[c]{@{}c@{}}\textbf{0.8806}\\(0.021) \end{tabular}                                 & \begin{tabular}[c]{@{}c@{}}0.8646\\ (0.021)\end{tabular}                   & \begin{tabular}[c]{@{}c@{}}\textbf{0.8845}\\ (0.048)\end{tabular} & \begin{tabular}[c]{@{}c@{}}\textbf{0.8879}\\ (0.018)\end{tabular} & \begin{tabular}[c]{@{}c@{}}0.8621\\ (0.019)\end{tabular}                   & \begin{tabular}[c]{@{}c@{}}\textbf{0.8873}\\ (0.049)\end{tabular} & \begin{tabular}[c]{@{}c@{}}\textbf{0.8906}\\ (0.02)\end{tabular}  \\ 
\hline
AALBERT        & \begin{tabular}[c]{@{}c@{}}0.8433\\ (0.021)\end{tabular}                  & \begin{tabular}[c]{@{}c@{}}0.8659\\ (0.055)\end{tabular}                  & \begin{tabular}[c]{@{}c@{}}0.871\\ (0.019)\end{tabular}  & \begin{tabular}[c]{@{}c@{}}0.8491\\ (0.022)\end{tabular}                   & \begin{tabular}[c]{@{}c@{}}0.8759\\ (0.052)\end{tabular}                   & \begin{tabular}[c]{@{}c@{}}0.8782\\ (0.019)\end{tabular} & \begin{tabular}[c]{@{}c@{}}0.8545\\ (0.023)\end{tabular}                   & \begin{tabular}[c]{@{}c@{}}0.8831\\ (0.051)\end{tabular} & \begin{tabular}[c]{@{}c@{}}0.8864\\ (0.02)\end{tabular}                    \\ 
\hline
Mockingjay     & \begin{tabular}[c]{@{}c@{}}0.8178\\ (0.024)\end{tabular}                  & \begin{tabular}[c]{@{}c@{}}0.852\\ (0.057)\end{tabular}                   & \begin{tabular}[c]{@{}c@{}}0.8582\\ (0.021)\end{tabular} & \begin{tabular}[c]{@{}c@{}}\textbf{0.8687}\\ (0.022)\end{tabular} & \begin{tabular}[c]{@{}c@{}}0.8624\\ (0.054)\end{tabular}                   & \begin{tabular}[c]{@{}c@{}}0.8687\\ (0.022)\end{tabular} & \begin{tabular}[c]{@{}c@{}}0.8312\\ (0.026)\end{tabular}                   & \begin{tabular}[c]{@{}c@{}}0.8719\\ (0.052)\end{tabular} & \begin{tabular}[c]{@{}c@{}}0.876\\ (0.021)\end{tabular}                    \\ 
\hline
wav2vec        & \begin{tabular}[c]{@{}c@{}}0.8498\\ (0.036)\end{tabular}                  & \begin{tabular}[c]{@{}c@{}}0.8691\\ (0.053)\end{tabular}                  & \begin{tabular}[c]{@{}c@{}}0.8738\\ (0.031)\end{tabular} & \begin{tabular}[c]{@{}c@{}}0.8348\\ (0.056)\end{tabular}                   & \begin{tabular}[c]{@{}c@{}}0.8736\\ (0.051)\end{tabular}                   & \begin{tabular}[c]{@{}c@{}}0.8777\\ (0.03)\end{tabular}  & \begin{tabular}[c]{@{}c@{}}0.8503\\ (0.037)\end{tabular}                   & \begin{tabular}[c]{@{}c@{}}0.877\\ (0.05)\end{tabular}   & \begin{tabular}[c]{@{}c@{}}0.8805\\ (0.028)\end{tabular}                   \\ 
\hline
APC            & \begin{tabular}[c]{@{}c@{}}0.8347\\ (0.064)\end{tabular}                  & \begin{tabular}[c]{@{}c@{}}0.8512\\ (0.05)\end{tabular}                   & \begin{tabular}[c]{@{}c@{}}0.8581\\ (0.061)\end{tabular} & \begin{tabular}[c]{@{}c@{}}0.84\\ (0.067)\end{tabular}                     & \begin{tabular}[c]{@{}c@{}}0.8616\\ (0.05)\end{tabular}                    & \begin{tabular}[c]{@{}c@{}}0.8663\\ (0.061)\end{tabular} & \begin{tabular}[c]{@{}c@{}}0.8467\\ (0.062)\end{tabular}                   & \begin{tabular}[c]{@{}c@{}}0.8626\\ (0.048)\end{tabular} & \begin{tabular}[c]{@{}c@{}}0.8695\\ (0.062)\end{tabular}                   \\ 
\hline
NPC            & \begin{tabular}[c]{@{}c@{}}0.8435\\ (0.032)\end{tabular}                  & \begin{tabular}[c]{@{}c@{}}0.8538\\ (0.054)\end{tabular}                  & \begin{tabular}[c]{@{}c@{}}0.8623\\ (0.032)\end{tabular} & \begin{tabular}[c]{@{}c@{}}0.8501\\ (0.033)\end{tabular}                   & \begin{tabular}[c]{@{}c@{}}0.8664\\ (0.053)\end{tabular}                   & \begin{tabular}[c]{@{}c@{}}0.8708\\ (0.032)\end{tabular} & \begin{tabular}[c]{@{}c@{}}0.8544\\ (0.033)\end{tabular}                   & \begin{tabular}[c]{@{}c@{}}0.872\\ (0.05)\end{tabular}   & \begin{tabular}[c]{@{}c@{}}0.8757\\ (0.031)\end{tabular}                   \\ 
\hline
DeCoAR         & \begin{tabular}[c]{@{}c@{}}0.8587\\ (0.056)\end{tabular}                  & \begin{tabular}[c]{@{}c@{}}0.8711\\ (0.049)\end{tabular}                  & \begin{tabular}[c]{@{}c@{}}0.8763\\ (0.054)\end{tabular} & \begin{tabular}[c]{@{}c@{}}0.8633\\ (0.054)\end{tabular}                   & \begin{tabular}[c]{@{}c@{}}0.8774\\ (0.047)\end{tabular}                   & \begin{tabular}[c]{@{}c@{}}0.8789\\ (0.05)\end{tabular}  & \begin{tabular}[c]{@{}c@{}}\textbf{0.8814}\\(0.019)\end{tabular} & \begin{tabular}[c]{@{}c@{}}0.8787\\ (0.046)\end{tabular} & \begin{tabular}[c]{@{}c@{}}0.8816\\ (0.054)\end{tabular}                   \\
\hline
\end{tabular}
\end{table*}

\subsection{Data pre-processing:}
The articulatory data corresponding to sensors attached to 6 articulators is passed through a low-pass filter to remove high-frequency noise incurred during the measurement process. The data is then normalized utterance level by mean removal and variance normalization. The recorded speech is downsampled to 16kHz before MFCCs and SSL features are extracted. MFCC is the signal-processing-based feature used since it is known to do well compared to other features such as LPC, LPCC, LSF etc, as shown in \cite{Ghosh2010AGS}. 13-dim MFCCs are extracted at 25ms windows with 10ms shift and EMA is downsampled to 100Hz so that it is aligned with MFCC. We choose SSL features which also produce features at the same frequency so that we don't have to further align these features. We found all SSL features to do well when the representation is extracted from the last layer of the SSL model, and hence we use the same.

\subsection{Model training and evaluation:}
We work on data from 10 subjects, each with 460 sentences. We use 80\% of data for training, 10\%     for validation and the remaining 10\% for testing. All sentences are unseen across sets and thus we evaluate on unseen sentence, seen speaker setup.  
We train the AAI model with 3 configurations, for all SSL features in 3 experimental setups - subject-specific (SS), pooled (P) and fine-tuned (FT). In subject-specific training, different models are trained for each subject, starting with default initialization. In pooled training, we combine the training data from 10 subjects and train a single model. For fine-tuning, we start with the pooled model weights and fine-tune each subject. 

Unlike \cite{udupa_aai_is21}, we don't arbitrarily zero pad to a fixed length, but rather apply padding based on the maximum sequence length in a single batch. We use a batch size of 16 for all experiments, with Adam optimizer and a learning rate of 0.0001. We use mean squared error loss over all the articulators, across the sequence as the loss function. A learning rate scheduler is used which lowers the loss when validation loss does not improve after a certain number of epochs. We use early stopping criteria based on validation loss to stop training the model. We use PyTorch to train our models, all codes are available at https://github.com/bloodraven66/ssl\_aai.

We obtain correlation coefficient \cite{Ghosh2010AGS} between the ground truth articulatory data and predicted trajectories, averaged over the test set utterances as well as the articulators, as our evaluation metric.

\label{sec:typestyle}

\section{Results and Discussion}
In this section, we interpret the results of AAI in different configurations. We first go over the effect of model size and then compare across various SSL features. Further, we look into the performance across different training setups of SS, pooled and FT and then observe the articulator-specific results.

\subsection{Effect of model size on features:}

Since input features widely vary in dimension, we experiment on 3 different model sizes with AAI-s, AAI-m, and AAI-l. From table \ref{tab:full_res}, we observe that in most cases, there is a slight increase in CC with an increase in model size from small to medium, and medium to large has little improvement. There are some exceptions, in the case of wav2vec where there was no improvement. Also, in the case of Mockingjay, the SS CC improved from 0.8178 to 0.8687. With FT, there has been a reduction in performance, for example, in TERA. These results suggest that the CC values increase with an increase in model size, with some exceptions in some features in certain scenarios.

\subsection{Investigating the performance of features:}

In this section, we compare the results across different SSL features and MFCC in table \ref{tab:full_res}. The first observation is that all SSL features except vq\_wav2vec perform better than MFCC. The drop in performance in vq\_wav2vec could be due to the quantization step in the SSL model. We also observe that there are mainly 3 SSL features - TERA, Mockingjay and DeCoAR which have obtained the best CC across configurations, with TERA having the best CC across 7 setups. While Mockingjay was the best in 1 configuration, we see that it doesn't do well in many other setups. Whereas, DeCoAR is close to TERA performance in many cases. On closer inspection of the features, SSL features which have a masked reconstruction objective function over log Mel spectrogram are generally working well. With TERA, we hypothesise that the masked and distortion-based reconstruction objective could be learning features representative of the log Mel spectrogram while being more generalizable, thanks to the robustness developed through the pretraining process. TERA is trained over masking over time and frequency domain, and corrupting frames by additive gaussian noise and randomly inserting frames. So these types of augmentations could also be beneficial to AAI if it is applied over standard acoustic features, and recently, there has been work \cite{siriwardena2022audio} towards this approach.

\subsection{Effect of SSL features on experimental setup:}
In this section, we look at the results w.r.t training setup - SS, pooled or FT from table \ref{tab:full_res}. We notice that there is a lot of improvement in CC in SS setup with SSL over MFCC. In medium and large models, SS results with TERA are better than the best results (FT) with MFCC. We can infer from this that in many training conditions, using SSL features on a small amount of data is better than using MFCCs on a much larger corpus (where SS and FT can be applied). We also observe that with the best FT SSL model, we can improve the AAI performance across all subjects to 0.89 CC when compared to the best of 0.8724 from MFCC. This indicates that even in the presence of more data, SSL features are more suited as input representation.

\subsection{Understanding articulatory specific variation:}
In this section, we interpret the results from table \ref{tab:art_spec} which reports the articulator results averaged across all subjects. We observe that there is improvement in CC in all articulators, especially for specific articulators such as $UL_y$, $LLy$, $Jaw_y$ etc in SS.





\label{sec:majhead}

\section{Conclusions}
In this work, we investigate the effectiveness of SSL features for AAI. We experiment with different types of SSL features, with different model sizes of transformer AAI and report across different experimental setups. We find that some of the SSL features based on masked reconstruction objectives can perform better than the standard MFCCs, even in single-speaker setups. This suggests that the SSL features have better utterance level generalizability, and these features could be an alternative to learning representation to predict articulatory movements. In the future, we will look at how SSL features can work with unseen speakers.

\footnotesize
\bibliographystyle{IEEEbib}
\bibliography{Template}

\end{document}